\documentclass[%
 amsmath,amssymb,
 aps,
 prf,
superscriptaddress,
longbibliography
]{revtex4-2}

\usepackage{dcolumn}
\usepackage{bm}

\usepackage{amssymb}
\usepackage{mathpazo}
\usepackage{siunitx}
\usepackage{amsmath}
\usepackage{pxfonts}

\usepackage{caption}
\usepackage{subcaption}
\usepackage{graphicx}
\usepackage{xcolor}

\usepackage{ulem}

\binoppenalty=\maxdimen
\relpenalty=\maxdimen

\definecolor{Cell2_db22}{rgb}{0, 0.55, 0.06}
\definecolor{Cell2_db33}{rgb}{0.69, 0.14, 0.69}
\definecolor{Cell3_db33}{rgb}{0.8672,    0.4922,    0.1211}

\begin{document}

\title{Bubble dynamics in an inclined Hele-Shaw cell}


\author{Benjamin Monnet}
\affiliation{ENS de Lyon, CNRS, Laboratoire de Physique, F-69342 Lyon, France.}

\author{J. John Soundar Jerome}
\affiliation{Univ Lyon, Universit\'{e} Claude Bernard Lyon $1$, Laboratoire de M\'{e}canique des Fluides et d'Acoustique, CNRS UMR--$5509$, Boulevard $11$ novembre $1918$, F--$69622$ Villeurbanne cedex, Lyon, France.}

\author{Val\'{e}rie Vidal}
\affiliation{ENS de Lyon, CNRS, Laboratoire de Physique, F-69342 Lyon, France.}

\author{Sylvain Joubaud}
\email{sylvain.joubaud@ens-lyon.fr}
\affiliation{ENS de Lyon, CNRS, Laboratoire de Physique, F-69342 Lyon, France.}

\begin{abstract}
We report experimental results on the dynamics of large bubbles in a Hele-Shaw cell subject to various inclination angles with respect to gravity. Low Reynolds number cases are studied by injecting bubbles in an stagnant water/UCON mixture in three different Hele-Shaw cell geometry. The leading order rise speed $v_b$ follows the Taylor-Saffman limit which is inversely proportional to the viscosity $\eta$, but directly proportional to the square of the cell gap $h$ and the effective gravity, accounting for cell tilt angle $\theta$. However, when the cell is inclined more and more, the bubble buoyancy in the cell gap leads to a substantial decrease in the rise speed, as compared to the Taylor-Saffman speed. Buoyancy pushes the bubble towards the top channel wall, whereby a difference between the lubrication film thickness on  top of and underneath the rising bubble occurs. We attribute these observations to the loss of symmetry in the channel gap, due to cell inclination. Nonetheless, the top lubrication film is observed to follow the Bretherton scaling, namely, $(\eta v_b/\sigma)^{2/3}$, where $\sigma$ is the liquid surface tension while the bottom film does not exhibit such a scaling. Finally, we illustrate that a model incorporating a friction term to the power balance between buoyancy and viscous dissipation matches well with all experimental data.
\end{abstract}

\maketitle

\section{Introduction}
\label{sec:intro}

Bubbly flows in a liquid or in a suspension involve many fascinating physics \cite{Lohse_2018, Dauxois_2021} and are widely present in nature \cite{Johnson_2002, Oppenheimer_2015, Vergniolle_2022} and industrial applications \cite{Kawchuk_2015, Du_2005}. 
Among these mutiphase flows, bubbles in a confined channel arise in numerous practical applications, like bioreactors, heat exchangers, gas absorption chambers, etc. In order to understand the complex flow situations in the presence of bubble swarms, it is necessary to obtain a thorough knowledge of the rising motion of a single bubble, and the bubble-induced flow in its surrounding. For this purpose, many studies were dedicated on the hydrodynamic behavior of a large bubble in a Hele-Shaw cell, i.e., a thin, wide rectangular channel. In the high Reynolds number regime, they focus on the planar oscillatory motion of a large bubble and its relation to shape deformation in a vertical cell, as for example in experiments by \citet{Kelley1997, Bush1997, Bush1998, Roig_2012, Filella_2015, pavlov_2021} and in numerical simulation of \citet{Wang_2014}. Whereas in the low Reynolds number regime, they study the dynamics of air bubbles driven by a viscous liquid within a horizontal Hele-Shaw channel, with \cite{gaillard2021, Keeler2022} or without \cite{Park_1984, Tanveer_1987, KopfSil_1988, Tanveer_1990} the presence of a centred depth perturbation. Following the work of \citet{Taylor_1959, Collins_1965} and \citet{Maxworthy_1986}, lately \citet{Monnet_2022} proposed a mechanistic model for the bubble rise speed when the Reynolds number is increased from viscous to inertial conditions in a vertical Hele-Shaw cell. In summary, these previous works focused mainly on two specific conditions for bubble dynamics, namely, due to buoyancy-induced or flow-imposed bubble motion in a vertical or horizontal Hele-Shaw, respectively.

The effect of non verticallity on the rise of a bubble has been explored in different confined environnements such as tubes \cite{Zukoski1966, Cavanagh_1999, Aussillous_2000, Shosho_2001, Massoud_2020} and near an inclined wall \cite{Aussillous2002, Podvin2008, Griggs2009, Dubois2016, Barbosa2019}. However, there are only a few recent works on quasi-planar bubble rise in a tilted Hele-Shaw cell. In particular, \citet{Tihon2019} experimentally explored the bubble rise velocity through an inclined, thin rectangular channel with strong lateral confinement and inertia effects. They illustrated that the classical large Reynolds number bubble speed $\sqrt{g \cos\theta \ell}$ \cite{Davies_1950, Collins_1965} remains valid in the case of inclined flat Hele-Shaw cells, if one accounts for the effective gravity $g \cos \theta$ due to cell inclination $\theta$ with respect to the vertical, and a characteristic length $\ell$ of the problem geometry. However, in the presence of a tilt angle, a rising bubble is generally not expected to stay along the center line of the cell gap. So, it is neither well-established nor straightforward whether an effective gravity correction is fully sufficient in the case of bubble motion in the viscous regime, at low Reynolds number. While such inhomogeneities along the cell gap coordinate might not be important in the inertial regime, it could strongly impact the viscous dissipation in a Hele-Shaw cell. This is the precisely the object of the present investigation. 

\begin{figure}
\centering 
      \includegraphics[]{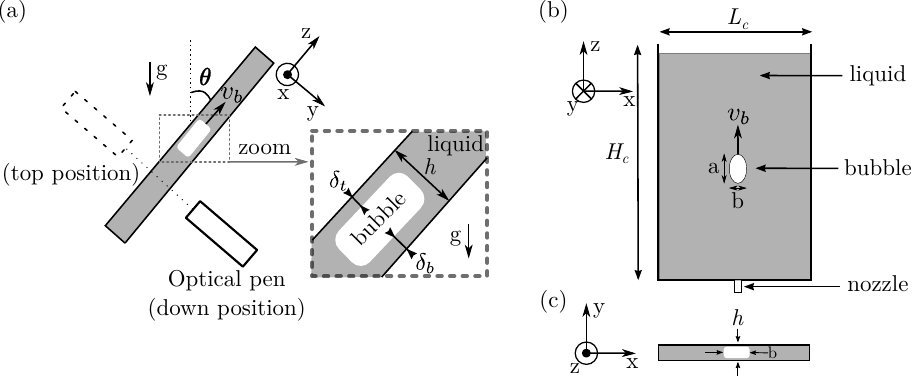}
  \caption{
  Sketch of the inclined Hele-Shaw cell with a single rising bubble in a liquid confined between two glass plates such that the gap $h$ is small compared to the cell height $H_c$ and width $L_c$ of the apparatus (not to scale). \textbf{(a)} Projection on the $yz$ plane. The CCS Prima OP 10000 optical pen locally measures the thickness of the lubrication film above or below the bubble (see section~\ref{sec:lubrication}).  \textbf{(b-c)} Projection on the $xz$ plane (b) and the $xy$ plane (c). $a$ and $b$ indicate the bubble size in the direction and perpendicularly to the movement.}
\label{fig:schema_exp}
\end{figure}

In this context, this work proposes to investigate the dynamics of a bubble rising in a tilted Hele-Shaw cell (Figure~\ref{fig:schema_exp}). The appropriate Reynolds number comparing inertial and viscous effects in this configuration is \cite{Batchelor_2000}
$$Re_{2h}=\dfrac{\rho v_b d_2}{\eta} \left(\dfrac{h}{d_2}\right)^2,$$
where $\rho$ is the liquid density, $\eta$ the liquid dynamic viscosity, $v_b$ the bubble speed and $d_2=2(A/\pi)^{1/2}$ the equivalent diameter computed from $A$, the area occupied by the bubble in the plane of the plates ($xz$). Taylor \& Saffman \cite{Taylor_1959} and Maxworthy \cite{Maxworthy_1986} proposed that, at low Reynolds number ($Re_{2h}\ll 1$), a very large elliptical bubble $d_2 \gg h$ rises at a constant speed 
\begin{equation}
    v_M=v_M^{\star} \frac{a}{b} \,\, ,  \,\,\,\, ~\text{with}~v_M^{\star} \equiv \frac{\Delta \rho (g \cos \theta) h^2}{12 \eta},
    \label{eq:Velocity_Maxworthy}
\end{equation}
where $a$ and $b$ are the length of the bubble in the direction and perpendicularly to the movement, respectively (see Figure~\ref{fig:schema_exp}b), $\Delta \rho=\rho-\rho_g$ with $\rho_g$ the density of the gas in the bubble, $g$ the gravitational acceleration and $\theta$ the tilt angle between the cell and the gravity (Figure~\ref{fig:schema_exp}a).

\begin{table}[h]
  \begin{center}
\begin{tabular}{cccccc}
\hline
\\
     Reference & Imposed &Angle  &Reynolds number  &Lubrication &Aspect  \\
     {}		&{flow}		&{$\theta$}		&{ $Re_{2h}$ }		&{film}		&{ratio}  \\
     {} & {}&{} &{}  &{} &{}  \\
\hline
     {} & {}&{} &{}  &{} &{}  \\
     Eck \& Siekmann, 1978 \cite{Eck_1978} & no & $80^{\circ}- 90^{\circ}$ & $10^{-5}-10^1$  & no & $\chi < 1$   \\ 
     Maxworthy, 1986 \cite{Maxworthy_1986}& no &$70^{\circ}-86^{\circ}$ & $10^{-3}-10^{-2}$ & no & $\chi > 1$ \\
     Filella et al., 2015 \cite{Filella_2015}& no & 0$^{\circ}$ & $10^1-10^{3}$  & no & $\chi < 1$ \\
     Monnet et al., 2022 \cite{Monnet_2022}& no & 0$^{\circ}$ & $10^{-4}-10^2$ & no & $\chi>1$ and $\chi<1$ \\ 
     Kopfsill \& Homsy, 1988 \cite{KopfSil_1988} & yes & $90^{\circ}$ & $10^{-6}-10^{-3}$  & no & - \\
     Gaillard et al., 2021 \cite{gaillard2021} & yes & 0$^{\circ}$ & $10^{-2}-10^0$ & yes & - \\
     This work & no & 0$^{\circ}-80^{\circ}$ & $10^{-3}-10^{-1}$ & yes & $\chi >1$\\
     {} & {}&{} &{}  &{} &{}  \\
       \hline
\end{tabular}
  \end{center}
    \caption{Existing experimental works on single bubble motion in Hele-Shaw cells filled with a Newtonian fluid. Note that, in the case of an imposed flow, a large variety of bubble shapes can be observed, including non-elliptical shapes as in \citet{KopfSil_1988}, and bubble splitting as in \citet{gaillard2021} when the cell gap is perturbed.}
    \label{table:papers}
\end{table}

In vertical cells ($\theta=0^\circ$), the theoretical velocity given by equation~\ref{eq:Velocity_Maxworthy} was experimentally recovered  \cite{Eri_2011, Madec_2020, Monnet_2022}. \citet{Eck_1978} and \citet{Maxworthy_1986} reported the first experimental data on large bubbles freely rising in tilted Hele-Shaw cells. However, the cells used in their investigations were often quasi-horizontal, with $\theta > 80^\circ$. \citet{Eck_1978} obtained steady rising of large single bubbles which are all oval but flattened in their direction of motion. On the other hand, \citet{Maxworthy_1986} reported large bubbles which are elongated along the rising direction. In the absence of any theory to predict the bubble shape, characterised by its aspect ratio $\chi=a/b$, so far \citet{Monnet_2022} observed that $\chi > 1$  in the viscous regime. Also, it does not significantly vary with any physical parameter, including the viscosity, the surface tension or the volume of the bubble. As of now, no other experimental evidence exists for oval bubbles which are flattened in the rising direction ($\chi < 1$) as in \cite{Eck_1978}. In other experimental conditions, i.e. at imposed flow rate, also close to the horizontal configuration, a wide variety of bubble shapes \cite{KopfSil_1988} with a larger range of aspect ratio was observed.

Lubrication films provide additional information on bubbles in a confined environment. Table~\ref{table:papers} presents previous experimental works on lubrication film of single bubble motion in Hele-Shaw cell. To our knowledge, no film thickness measurement has ever been performed for bubbles in a vertical or inclined Hele-Shaw cell in the absence of counterflow. Note that previous works have quantified the lubrication films for falling drops in the same geometry \citep{Keiser_2018, Reichert_2019, Toupoint_2021}. The scope of the present work is to focus in particular on freely rising bubbles in an inclined Hele-Shaw cell (Table~\ref{table:papers}). \citet{Park_1984} developed a theoretical analysis for the lubrication films, based on the Bretherton analysis \cite{Bretherton_1961}, in the case of a quasi-horizontal configuration with an imposed flow. From this perspective, film thickness measurement could not only help to fill the gap in the existing literature, but also provide founding stones for future theoretical work.

In the midst of these previous results, the present paper aims at a systematic study of a bubble rise in an inclined Hele-Shaw cell, in the low  Reynolds number regime ($Re_{2h} \ll 1$). In particular, we will characterize the bubble velocity and shape, as well as the lubrication film thickness, as a function of the inclination angle, cell gap, liquid viscosity and bubble size.


\section{Experimental setup}
\label{sec:expsetup}

The experimental setup consists of a Hele-Shaw cell made of two glass plates of height $H_c$ and width $L_c$ separated by a thin gap $h$, which can be inclined by an angle $\theta$ between $0^\circ$ and $80^\circ$ respect to the vertical (Figure~\ref{fig:schema_exp}). Table~\ref{table:cells} lists the physical properties of the three different cells and various liquids used in the present study.
The width-to-length ratio is chosen in such a way that lateral confinement is negligible in experiments reported here \cite{Gondret_1997, pavlov_2021}. Cell 1 and Cell 2 are comparable in size and gap, but the glass plate thickness and the gas injection system are slightly different. For a given cell at a fixed inclination angle $\theta$, $h$ is approximately uniform. However, measurements using an optical sensor, and/or by filling the cell with a known quantity of water indicate a slight dependence of the cell gap $h$ with the tilt angle $\theta$, up to 3~\% (see Appendix~\ref{sec:htheta}). The bubble terminal velocity predicted in the viscous regime varies as $h^2$ (Eq.~\ref{eq:Velocity_Maxworthy}) and a slight variation of the cell gap may play a significant role. Therefore, for each experiment, we have carefully measured $h(\theta)$ and systematically used the experimental value for each angle. In the following, $h_0$ indicates the cell gap at $\theta=0^\circ$ (vertical cell).

\begin{table}[h!]
  \begin{center}
  \begin{tabular}{ccccccccccc}
\hline
\\
        & $H_c$ ~ & ~$L_c$~  &~ $h_0 = h(\theta=0^{\circ})$ ~&~$e_g$ ~ & ~$\eta$~& ~$\gamma$~  & ~$\rho$~  & ~$Re_{2h}$~ & ~$\theta$~  &Symbol  \\ 
       & (cm) & (cm) & (mm) & (mm) & (mPa.s) & (mN.m$^{-1}$) & (kg.m$^{-3}$) &{(max)} & ($^{\circ}$) & \\ 
       &{} &{} &{} &{} &{} &{} &{} &{} &{} &{}\\ 
       \hline
       &{} &{} &{} &{} &{} &{} &{} &{} &{} &{}\\ 
       Cell 1 & 28.5 & 23.5 & 2.19 & 3.9  & $\left\{ 145,219 \right\}$& $\left\{ 51,51 \right\}$&$\left\{ 1048, 1055 \right\}$ & 0.10 & [0-75] &$\circ$  \\
       Cell 2 & 30 & 20 & 2.29 & 5 & 145&  51 &1048 &  0.12 & [0-80] &$\vartriangleright$ \\
       Cell 3 & 30 & 20 & 5.21 & 5 & 580&
       47 & 1068 & 0.17 & [0-80] & $\lozenge$\\
       &{} &{} &{} &{} &{} &{} &{} &{} &{} &{}\\
       \hline 
  \end{tabular}
  \end{center}
  \caption{Cell and liquid properties. Three different cells of height $H_c$, length $L_c$ and gap $h_0= h(\theta=0^{\circ})$ are used in this work. $e_g$ is the glass plates thickness. $\eta$, $\gamma$ and $\rho$ are the liquid viscosity, surface tension and density, respectively. For each cell, the maximum Reynolds number and the range of inclination angle $\theta$ explored in the experiments are given.}
    \label{table:cells}
\end{table}

All experiments are performed at room temperature. In our setup, water-\textsc{Ucon} mixtures are used to vary the liquid viscosity $\eta$. 
The viscosity of these mixtures depends on temperature, and the ambient temperature can slightly vary. Therefore, for each experiment, the room temperature was measured and we determined the associated fluid viscosity at the same temperature($\pm~0.5^{\circ}$C) by means of a Kinexus Ultra+ rheometer. For sake of clarity, throughout the manuscript all viscosity values are given at $20^{\circ}$C, but the temperature dependence was properly accounted for in the calculations that involve the liquid viscosity $\eta$. The surface tension $\gamma$ and liquid density $\rho$ are quantified using the pendant drop method with an \textsc{Attension Theta} tensiometer (Biolin Scientific) and an \textsc{Anton~Paar~DMA~35} densimeter, respectively.


Bubbles are generated at the centre of the cell's bottom with the help of a millimetric-sized pipe attached to a manually-controlled $50$~mL syringe.
For each run, the cell is backlit uniformly with a LED panel while a computer-controlled camera (Basler acA2440, $2048\times1024$ pixels) records the rising motion of the bubble at $5$ to $30$ fps, depending on the bubble velocity. Note that although we are able to visualize the whole cell, the bubble motion is analyzed in a region of interest far from the cell boundaries ($\approx 4$ cm from the top and bottom, and 8 cm from the sides as the bubble roughly rise vertically). This ensures that the bubble is in the stationary regime and rises at constant speed. Images are then binarized (the threshold of binarization induces an error of less than $1$\% on the bubble characteristics) and standard techniques in \textsc{Matlab}$\circledR$ are performed to identify the bubble contour, define the equivalent ellipse and compute the bubble speed $v_b$ and aspect ratio $\chi=a/b$. We remind that $a$ and $b$ are the bubble axes parallel (longitudinal) and perpendicular (transverse) to its motion, respectively. The equivalent planar bubble diameter is therefore $d_2=\sqrt{ab}$. In the following, only bubbles with $d_2>h$ are investigated in order to stay in the quasi-2D approximation wherein the Taylor-Saffman bubble speed (Eq.~\ref{eq:Velocity_Maxworthy}) is valid. 


\section{Experimental results}
\label{sec:results}

\subsection{General observations}
\label{sec:observations}

\begin{figure}[h!]
\centering
      \includegraphics[]{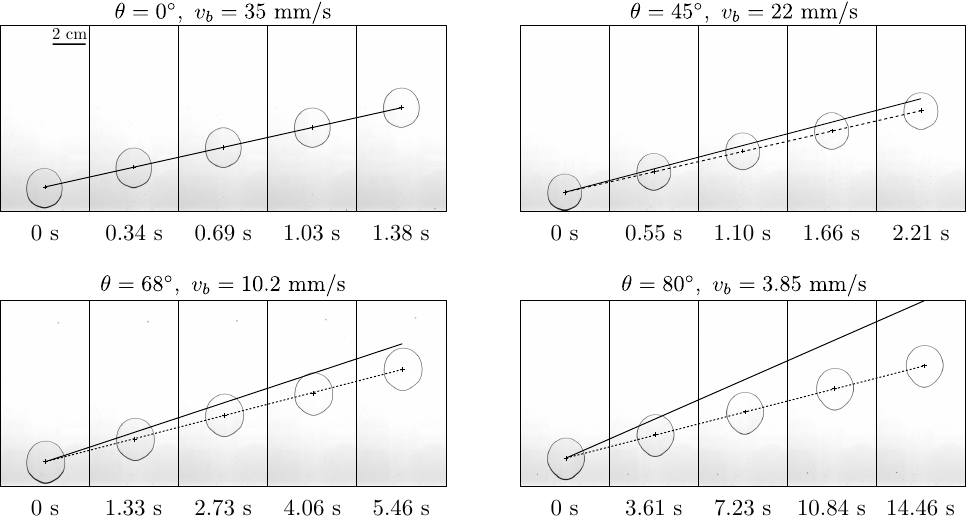}
\caption{Chronophotographies of four distinct bubbles rising in the same cell (Cell~1, $h_0 = 2.19$~mm, see Table \ref{table:cells}). Their equivalent diameter is identical ($d_2= 22 \pm 1$~mm) and the tilt angle $\theta$ is varied from $0^\circ$ to $80^\circ$. Dashed lines represent the observed bubble trajectories. Note that they are all straight lines, corresponding to a constant rise speed $v_b$. Solid lines indicate the trajectories estimated from the theoretical bubble speed (Eq.~\ref{eq:Velocity_Maxworthy}) considering the effective gravity $g \cos \theta$ (see section~\ref{section:bubble_speed} for more details). The discrepancy with the real trajectory increases with the tilt angle. All images are at the same scale.}
\label{fig:chrono_v1}
\end{figure}

Figure~\ref{fig:chrono_v1} shows four chonophotographies of individual bubbles rising in Cell 1 (see Table~\ref{table:cells}) for different inclination angle $\theta$. Here, the bubble equivalent diameter ($d_2 = 22 \pm 1$~mm) and viscosity $(\eta=145~$mPa.s) are kept identical for sake of comparison.  Dashed lines correspond to the trajectory of each bubble's center of mass. For a given bubble, we notice that the speed and the bubble shape remain constant. No shape or velocity oscillations were observed in our experiments as it is often the case in the inertial regime~\cite{Filella_2015}. This observation is in agreement with the fact that the viscous time scale $\tau_{\eta}=h^2\rho/(4\eta)$~\cite{Filella_2015}, which is lower than $1$~ms in our study, is small compared to the rising time of the bubble or the time between two images. So, in what follows, we consider only time-averaged values of the bubble speed and aspect ratio. Errorbars are provided to show the standard deviation from such averages. Figure~\ref{fig:chrono_v1} illustrates that the rise speed of the bubble decreases monotonically from 35~mm/s to 3.85~mm/s when the tilt angle $\theta$ is increased from 0$^{\circ}$ to 80$^{\circ}$. In all our experiments, a typical bubble attains a terminal velocity and displays an elongated oval shape along the rising direction, characteristic of the viscous regime. As expected, its rise speed decreases when the Hele-Shaw cell is tilted towards the horizontal axis. Interestingly, although the theoretical bubble speed considering the effective gravity $g \cos \theta$ (Eq.~\ref{eq:Velocity_Maxworthy}) predicts well the bubble dynamics in a vertical cell (Figure~\ref{fig:chrono_v1}, $\theta=0^\circ$), it overestimates the bubble velocity when inclining the cell (solid lines, Figure~\ref{fig:chrono_v1}), and this discrepancy increases when increasing $\theta$.


\subsection{Bubble aspect ratio and speed}
\label{section:bubble_speed}

In order to compare its rise speed to the theoretical value given by Eq.~\ref{eq:Velocity_Maxworthy}, it is necessary to measure the bubble aspect ratio. In this section we first report and comment the bubble aspect ratio $\chi$ and then present the evolution of the time-averaged bubble speed with the tilt angle $\theta$. 

\begin{figure}[h!]
\centering
      \includegraphics[]{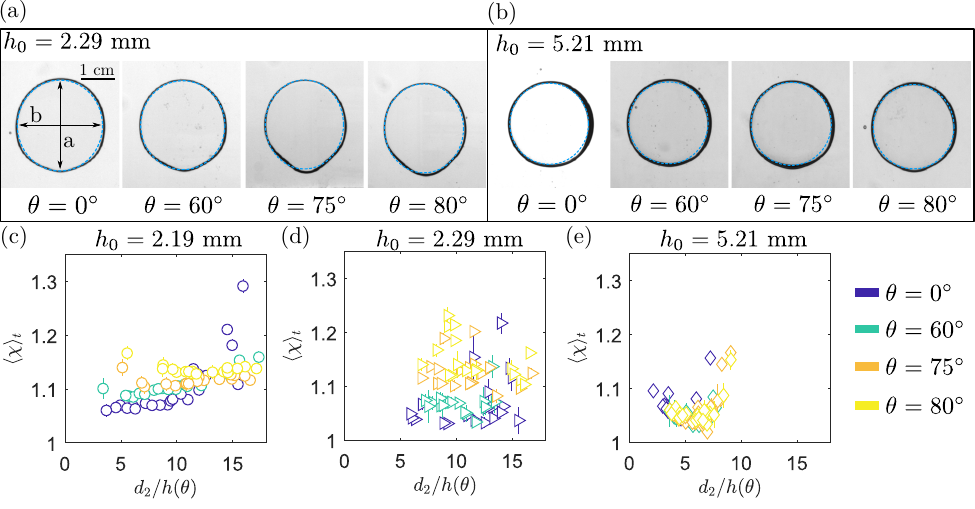}
  \caption{    \textbf{(a)}--\textbf{(b)} Instantaneous photos of bubbles with the same size $d_2=25 \pm 1$~mm for various tilt angles $\theta$. The viscosity is varied from (a) to (b) to keep the Reynolds number smaller than 1 ($Re_{2h}\ll 1$).  \textbf{(a)}  Cell 2 ($h_0=2.29$~mm), $\eta=145$~mPa.s); \textbf{(b)} Cell 3 ($h_0=5.21$~mm), $\eta=580$~mPa.s). Dashed lines (blue) correspond to an elliptical fit over the bubble contour. All the images have the same scale. \textbf{(c)}--\textbf{(e)} Time-averaged aspect ratio of bubbles against their normalised diameter $d_2/h(\theta)$ for $\theta=\{0,60,75,80\}^{\circ}$ in Cell 1 (c), Cell 2 (d) and Cell 3 (e).}
\label{fig:RA_all_h}
\end{figure}

Figure~\ref{fig:RA_all_h}(a) shows instantaneous photographs of bubbles in Cell 2 ($h_0 = 2.29$~mm) for a bubble diameter $d_2=25 \pm 1~$mm at four different cell tilt angles $\theta=\{0^\circ,60^\circ,75^\circ,80^\circ\}$. A small bulge is observed at the rear end of the bubbles for the highest $\theta$ values. Figure~\ref{fig:RA_all_h}(b) displays bubbles with the same apparent diameter $d_2$ in Cell~3 with a larger gap ($h_0 = 5.21$~mm), for the same inclination angles than Figure~\ref{fig:RA_all_h}(a). Note here that the viscosity has been changed so that the Reynolds number $Re_{2h}$ in both cells is comparable for a given angle $\theta$, for sake of comparison. In the cell of larger gap, the bulge is less pronounced and the aspect ratio remains approximately constant over different tilt angles (Figure~\ref{fig:RA_all_h}(b)). Although at present the bulge origin cannot be explained, we can see from experimental observations that the bubble contour is well-fitted by an elliptical shape (dashed blue lines, Figures~\ref{fig:RA_all_h}(a)--(b)). Those results are consistent with most of the previous works~\cite{Maxworthy_1986, Madec_2020, Monnet_2022}, which report elongated bubbles in the direction of motion ($\chi >1$) in the viscous regime without imposed flow (see Table~\ref{table:papers}). Note that  Eck \& Siekmann~\cite{Eck_1978} observed flattened bubbles ($\chi < 1$) in water/isopropanol mixtures. Figures~\ref{fig:RA_all_h}(c)--(e) display the evolution of the time-averaged aspect ratio $\langle \chi \rangle_t=\langle a/b \rangle_t$ against the normalised diameter $d_2/h(\theta)$ for Cells 1, 2 and 3, respectively. The choice of $d_2/h$ has been made to distinguish between bubbles which are fully in the 2D approximation ($d_2 \gg h$) and bubbles for which 3D effects are not negligible anymore ($d_2 \sim h$). Here, colors indicate different tilt angles $\theta=\{0,60,75,80\}^{\circ}$. At a given tilt angle $\theta$, the aspect ratio is almost the same for each bubble. This implies that $\langle \chi \rangle_t$ only weakly depends on the equivalent bubble diameter $d_2$, except for larger bubbles ($d_2/h>15$ for Cell 1 and $d_2/h>8$ for Cell 3) for which the values of $\langle \chi (t) \rangle_t$ are higher. This augmentation has already been obersved by \citet{Madec_2021}.  However, as inferred previously from the photographs, the aspect ratio seems to slightly increase with $\theta$ for the smallest gaps $h_0=2.19$~mm and $h_0=2.29$~mm, while no dependence on the angle is observed for $h_0=5.21$~mm in our experiments. Note that the data, especially for $h_0=2.29~$mm (figure 3d), are scattered, which perhaps arises from a less controlled injection system. Nonetheless, despite the scattering, the evolution of $\langle \chi \rangle_t$ with $\theta$ is statistically reproducible.

\begin{figure}[h!]
\centering
      \includegraphics[width=0.9\linewidth]{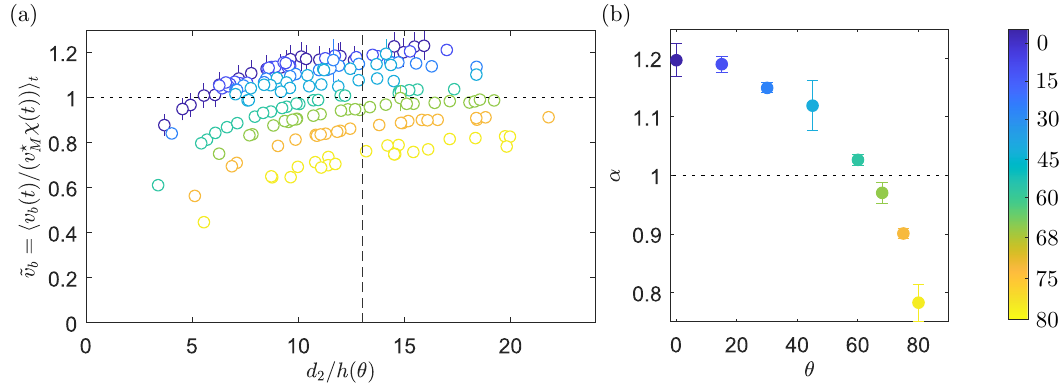}
  \caption{\textbf{(a)} Evolution of the non-dimensional bubble velocity $\tilde{v}_b =\langle v_b(t) / v_M(t) \rangle_t$, where $v_M(t) = v_M^{\star} \chi(t)$, with $d_2/h(\theta)$ for different inclination angles $\theta$ (see colorbar) [Cell 1, $h_0=2.19$~mm, $\eta=$145~mPa.s]. The maximum value of $Re_{2h}$ reached is $0.10$. The vertical dashed line indicates $d_2/h(\theta)=N=13$, the threshold used to define the plateau value $\alpha$ (see text). The horizontal dashed line indicates the theoretical value $\tilde{v}_b=1$. \textbf{(b)} Plateau value $\alpha$ defined by equation~\ref{eq:alpha_def} as a function of the tilt angle $\theta$. The black dashed line indicates the theoretical value $\alpha=1$. The errorbars represent the standard deviation of the values used to compute $\alpha$.}
\label{fig:VbSurVm_angles}
\end{figure}

Figure~\ref{fig:VbSurVm_angles}(a) displays the time-averaged bubble rise speed as the bubble size is increased towards the theoretical limit $d_2 \gg h$. All data correspond to experiments in Cell 1 ($h_0=2.19$ mm, $\eta=145~$mPa.s). For each data point, the normalised velocity $\tilde{v}_b$ is computed via a time-average of the ratio of the instantaneous bubble speed over the Taylor-Saffman speed i.e., $\tilde{v}_b = \langle v_b(t) / v_M(t) \rangle_t$, where $v_M(t) = v_M^{\star} \chi(t)$ remains almost constant as the variations of $\chi (t)$ are small, as seen in part~\ref{sec:observations}. For a given tilt angle $\theta$, the non-dimensional speed $\tilde{v}_b$ increases with the normalised bubble size $d_2/h(\theta)$. Furthermore, measurements strongly suggest that $\tilde{v}_b$ tends towards a plateau value $\alpha$ corresponding to the limit when $d_2/h(\theta) \gg 1$. These observations are valid for each cell inclination, with the plateau value depending on the tilt angle, $\alpha=\alpha(\theta)$. The plateau value is defined as the mean normalised bubble speed $\tilde{v}_b$ for all bubbles which satisfy $d_2/h(\theta) > N$ .i.e,
\begin{equation}
    \alpha(\theta) =\langle \tilde{v}_b \rangle_{d_2/h(\theta)>N},
    \label{eq:alpha_def}
\end{equation}
where $N$ is chosen arbitrarily large and independent of $\theta$, here $N=13$. The variation of the maximum normalised rise speed $\alpha$ with tilt angle $\theta$ is presented in Figure~\ref{fig:VbSurVm_angles}(b). The plateau value $\alpha$ decreases steadily with the inclination angle, despite the explicit dependence of the Taylor-Saffman speed on $\theta$ via the effective gravity $g \cos \theta$. It results in a relative variation of $\alpha$ of around 30$\%$ between 0$^{\circ}$ and 80$^{\circ}$. In other words, the bubble speed given by Eq.~\ref{eq:Velocity_Maxworthy} does not predict accurately the dependence of the bubbles velocity with $\theta$. Note that the theoretical rise speed is not recovered for the vertical configuration i.e, $\alpha(\theta=0^\circ) \equiv \alpha_0=1.20\pm0.03\neq 1$. We attribute this observation to the fact that the cell gap $h$ and liquid viscosity $\eta$ are precisely measured in the present setup. Nonetheless, this result does not change the conclusions of previous articles on the impact of the Reynolds number $Re_{2h}$, or the presence of grains in the vertical configuration~\cite{Madec_2020, Monnet_2022}.  As this work focuses on the dependence on the inclination angle $\theta$, in the following we will consider the normalised plateau value, $\alpha/\alpha_0 = \alpha(\theta)/\alpha(\theta=0^\circ)$. Note that the value of $\alpha_0$ is not the same for the three cells but it depends neither on $\eta$ nor $h$.

\begin{figure}[h!]
\centering
      \includegraphics[width=0.9\linewidth]{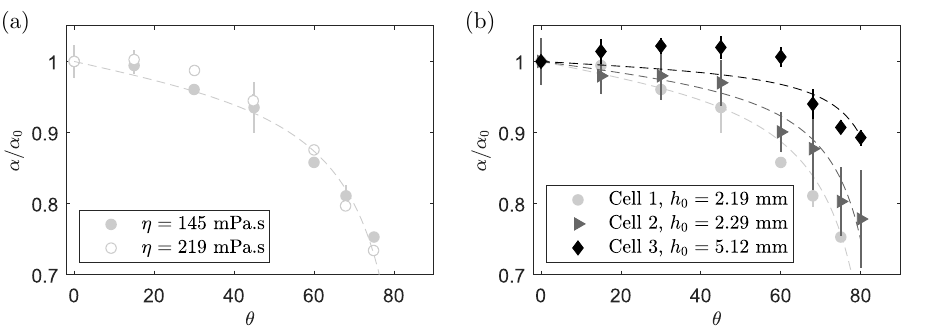}
  \caption{normalised plateau value $\alpha/\alpha_0= \alpha(\theta)/\alpha(\theta=0^\circ)$ as a function of the cell inclination angle $\theta$ \textbf{(a)} for two different viscosities (Cell 1, $h_0=2.19$~mm), \textbf{(b)} for different cell gap $h_0$. Dashed lines correspond to $\alpha/\alpha_0=(1-\kappa \tan \theta)$ (Eq.~\ref{eq:modele_alpha}), where $\kappa$ is obtained from best fit: $\kappa=(6.5 \pm 0.5)~\times10^{-3}$ for Cell 1, $\kappa=(4.5 \pm 1.5)~\times10^{-3}$ for Cell 2, and $\kappa=(1.9 \pm 0.4)~\times10^{-3}$ for Cell 3.}
\label{fig:alpha_theta}
\end{figure}

Figure~\ref{fig:alpha_theta}(a) displays $\alpha/\alpha_0$ as a function of $\theta$ for two water/\textsc{Ucon} mixtures of different viscosity ($\eta=$~145~mPa.s and $\eta=$~219~mPa.s). Within the error bars, $\alpha/\alpha_0$ does not depend significantly on the fluid viscosity. Figure~\ref{fig:alpha_theta}(b) compares the variations of $\alpha/\alpha_0$ in Cell 1 ($\circ$) to different cell geometry ($\vartriangleright$ Cell 2, $h_0=2.29$~mm and $\lozenge$ Cell 3, $h_0=5.21$~mm). In contrast to the liquid viscosity, the variation of the plateau values between different cells are readily discerned at a fixed $\theta$. Such differences are further amplified in the limit when $\theta$ tends towards $90^\circ$. Thus, all data confirm the dependence of the maximum speed limit on the gap $h$ and the tilt angle $\theta$ when a large bubble rises in a tilted cell. By definition $\alpha = \tilde{v}_b/v_M$, wherein $v_M^{\star} \propto g \cos \theta h^2$ \cite{Eck_1978, Maxworthy_1986, Madec_2020, Monnet_2022} and so, these results cannot be accounted by the Taylor-Saffman limit (Eq.~\ref{eq:Velocity_Maxworthy}). However, the normalised maximum speed $\alpha_0$ for large bubbles $d_2 \gg h_0$ in a vertical Hele-Shaw cell is independent of cell gap $h_0$ and viscosity $\eta$. It can be expected that the symmetry about the $y$-axis is not maintained anymore when a bubble rises in an inclined Hele-Shaw cell. This could lead to the observed dependence of the bubble rise speed on the cell gap $h$ and tilt angle $\theta$. The next section is dedicated to the quantification of this asymmetry, via lubrication films measurements.

\subsection{Lubrication films}
\label{sec:lubrication}

\begin{figure}[t]
\centering
      \includegraphics[width=0.8\linewidth]{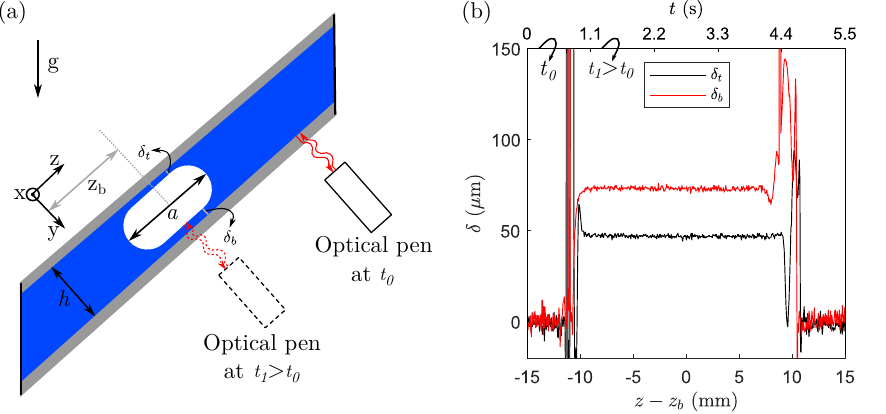}
 \caption{\textbf{(a)} Sketch in the frame of the rising bubble center of mass ($z=z_b$), in the $yz$ plane. The CCS optical pen is fixed in the laboratory frame, here located at the cell bottom (see Figure~\ref{fig:schema_exp}a). It initially measures the glass wall thickness only ($t=t_0$). When the bubble rises in front of the pen ($t=t_1$), it measures both the glass plate thickness and the lubrication film (times $t_0$ and $t_1$ are given as examples). $\delta_t$ and $\delta_b$ indicate the top and bottom lubrication film thickness, respectively.  \textbf{(b)} Typical signals of the CCS optical pen for $\delta_t$ and $\delta_b$ as a function of time $t$ (upper axis) or $z-z_b$ (lower axis), with $z_b$ the $z-$coordinate of the bubble center. The signals are obtained by subtracting the glass thickness for two different bubbles with the same apparent diameter ($d_2 = 22 \pm 1$~mm, Cell 2, $\theta=80^{\circ}$, $\eta=145$~mPa.s).}
\label{fig:signal_crayon_top_bottom}
\end{figure}

In the previous section, we have considered the bubble shape in the plane $xz$ of the cell plates. In this section, we will discuss the bubble shape in the gap i.e., in the $yz$ plane (Figure~\ref{fig:schema_exp}).
For this purpose, the lubrication film on the upper side and on the bottom side of a rising bubble has been measured by a Chromatic Confocal Sensing (CCS) Prima OP 10000 optical pen (STIL) located above or below the cell, respectively (Figure~\ref{fig:schema_exp}(a)). The CCS optical pen is  designed to send light at different wavelengths at well-defined time intervals. It then measures the distance to an interface by identifying the wavelength (color) of the reflected signal. Since the rising bubble is in steady motion with respect to the laboratory frame where the CCS optical pen is fixed, the pen initially measures the glass plate thickness, at say $t = t_0$, by analysing the light reflected at the liquid/glass interface (see Figure~\ref{fig:signal_crayon_top_bottom}a for illustration in the frame of the bubble). When a bubble intercepts the light sent by the optical pen, say at $t_1 > t_0$, the presence of air/liquid interface is instantaneously detected and the CCS optical pen measurement is accordingly modified. By properly taking the difference between these two outputs from the CCS optical pen, it is possible to compute the liquid film thickness between the bubble wall and glass plate. In the following, the top and bottom lubrication film thicknesses are referred to as $\delta_t$ and $\delta_b$, respectively. Note that no simultaneous measurement of $\delta_t$ and $\delta_b$ is possible with a single optical pen. However, the measurements are reproducible, and it is easy to produce bubbles with identical diameter $d_2$, which make possible to investigate the joint behavior of the top and bottom lubrication film as a function of the other experimental parameters.

Two typical signals from the CCS optical pen, one above ($\delta_t$, black line) and the other below ($\delta_b$, red line) are displayed in Figure~\ref{fig:signal_crayon_top_bottom}(b). As stated above, these signals were collected from two different trials with the same bubble diameter $d_2 = 20 \pm 1$~mm in the same experimental configuration (Cell 2). They are plotted both as a function of time $t$ (upper axis) and of $z-z_b$ (lower axis), with $z_b$ the bubble center so that the signal falls to zero when $\vert z_b - z \vert > d_2/2 = 11$~mm. At the edges of the bubble, when $\vert z_b - z \vert \approx 11$~mm, the signal from the pen is noisy because of the curvature of the air-liquid interface. Otherwise, the thickness measured by the pen is approximately constant across the bubble size $\vert z_b - z \vert < d_2/2$, apart from a small electrical noise.

\begin{figure}[h!]
\centering
      \includegraphics[width=1\linewidth]{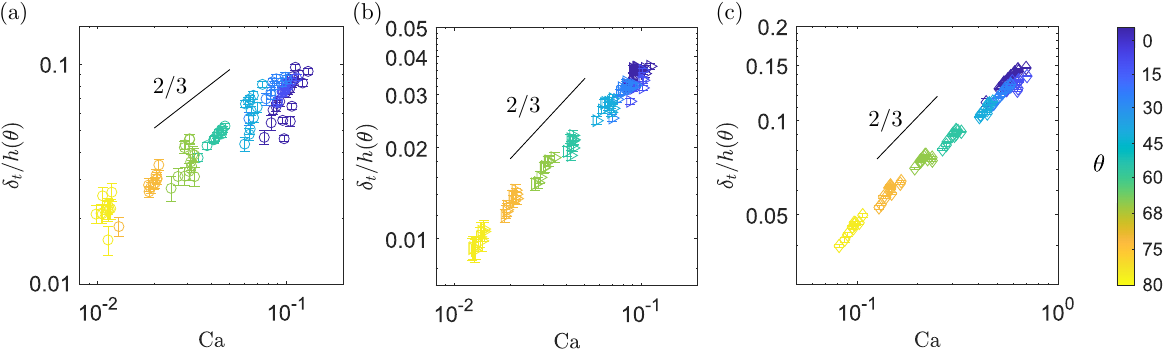}
  \caption{Non-dimensional top lubrication film thickness $\delta_t/h(\theta)$ as a function of the capillary number Ca$ = \eta v_b/\gamma$ for different inclination angles $\theta$ (see colorbar). \textbf{(a)}  Cell 1 ($h_0=2.19~$mm), \textbf{(b)} Cell 2 ($h_0=2.29~$mm) and \textbf{(c)} Cell 3 ($h_0=5.21~$mm).}
\label{fig:Lubrification_film_top}
\end{figure}

\citet{Bretherton_1961} was the first to propose a theoretical expression for the thickness of the lubrication film formed by a bubble rising in a cylindrical tube. \citet{Park_1984} later extended this study to bubbles in a Hele-Shaw cell with an imposed flow, showing that the lubrication films should scale as $\delta_b/h(\theta)=\delta_t/h(\theta)=1.337~\text{Ca}^{2/3}$ for Ca$\ll 1$, where $\text{Ca}=\eta v_b/\gamma$ is the capillary number. In a recent work by \citet{gaillard2021} in a horizontal Hele-Shaw cell with an imposed flow, by assuming the top and bottom lubrication film equal, the above results were experimentally recovered and extended to all Ca with the empirical relation
\begin{eqnarray}
\dfrac{\delta_b}{h_0} &= \dfrac{\delta_t}{h_0} &= \dfrac{c_1\text{Ca}^{2/3}}{1+c_1 c_2\text{Ca}^{2/3}},
\end{eqnarray}
where $c_1$ and $c_2$ are constants and estimated with best data fit. Based on these previous investigations, it is convenient to present the evolution of top film thickness $\delta_t/h(\theta)$ against the capillary number $\text{Ca}=\eta v_b/\gamma$.  Figures~\ref{fig:Lubrification_film_top}(a)--(c) presents such graphs for Cell~1 ($h_0=2.19~$mm), Cell~2 ($h_0=2.29~$mm) and Cell~3 ($h_0=5.21~$mm), respectively. In each figure i.e., at a fixed cell geometry, data over various tilt angles fall on a straight line in the log-log plot. Therefore, for inclined Hele-Shaw cells the ratio $\delta_t/h (\theta)$ also seems to scale well with Ca$^{2/3}$. However, a closer look shows that the range of the $y$-axis varies from one cell to the other. In addition, the prefactor of the Ca$^{2/3}$ scaling clearly depends on the cell characteristics. 

We observed no such trivial scaling of $\delta_b/h(\theta)$ with Ca$^{2/3}$ (see Appendix~\ref{sec:deltab}) for the lubrication film underneath the bubble. In fact, the symmetry in the $yz$ plane is broken as soon as the cell is tilted, and there is no reason that the lubrication film on the top and bottom of the bubble remain identical. Therefore, the capillary number is not sufficient to characterize the bottom film thickness. Such an observation also suggests that the Bretherton problem for at least one of the lubrication film of a rising bubble between rigid walls cannot be perhaps simplified to an analogous Landau-Levich drag-out problem in the frame of reference of a rising bubble.

\begin{figure}[h!]
\centering
      \includegraphics[width=0.75\linewidth]{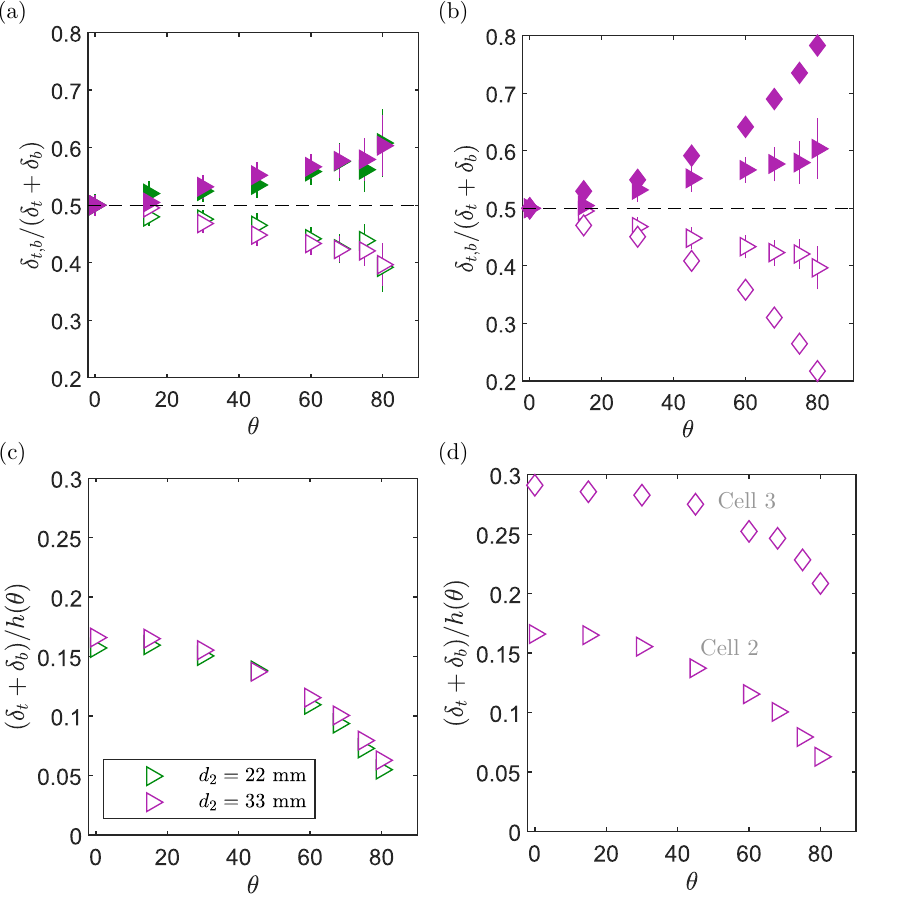}
      \caption{Ratio of the top ($\delta_t$, open symbols) and bottom ($\delta_b$, filled symbols) film thickness to the total film thickness $(\delta_b+\delta_t)$ as a function of tilt angle $\theta$ \textbf{(a)} in Cell 2, for two bubble sizes $d_2=22~$mm (green) and $d_2=33~$mm (purple); \textbf{(b)} for $d_2=33~$mm in Cell 2 ($\vartriangleright$, $h_0=2.29~$mm) and Cell 3 ($\lozenge$, $h_0=5.21~$mm). \textbf{(c)} Variation of the total lubrication film thickness $(\delta_t+\delta_b)$ by the cell gap $h(\theta)$ as a function of the inclination angle for two different bubble size $d_2$ (Cell 2). \textbf{(d)} $(\delta_t+\delta_b)/h$ in Cell 2 and 3 (bubble size $d_2=33~$mm).}
\label{fig:delta_rel_tot}
\end{figure}

To quantify the symmetry breaking in the $yz$ plane, we represent in Figure~\ref{fig:delta_rel_tot}(a) the ratios $\delta_t/(\delta_t+\delta_b)$ and $\delta_b/(\delta_t+\delta_b)$ as a function of various tilt angles, for two different bubble diameters ($d_2 = 22 \pm 1$~mm and $33\pm1$~mm, Cell 2). The relative contribution of the bottom lubrication film $\delta_b$ (filled symbols) to the total film thickness ($\delta_t + \delta_b$) increases as the inclination angle $\theta$ increases. This evolution is irrespective of the bubble size. Figure~\ref{fig:delta_rel_tot}(b) compares the same parameters, namely, $\delta_t/(\delta_t+\delta_b)$ and $\delta_b/(\delta_t+\delta_b)$, for two different cells (Cell 2 and 3). Here, the bubble equivalent diameter is fixed ($d_2=33\pm1$~mm). The relative size of the bottom lubrication film, as compared to the top one, is always more important in Cell 3 where the cell gap is at least twice larger than for Cell 2. We also report that the gap effect is amplified when the cell tilt angle $\theta$ is increased. Note that these trends are similar to the observations on the bubble rise speed in the previous section. So, we can conclude similarly that the cell inclination $\theta$ accentuates the influence of gravity perpendicular to the cell walls since $\sin \theta$ increases. Therefore, a rising bubble in a tilted cell experiences a stronger push towards the upper wall. It is then expected that such an effect could draw partially the liquid into the bottom lubrication film and explain the observed behaviour. In a larger cell, the quantity of air per surface unit (i.e., the pressure applied on the liquid film) that tries to go up is more important, resulting in an even larger difference between the liquid films.


Figure~\ref{fig:delta_rel_tot}(c) shows the evolution of the normalised total lubrication film $(\delta_t+\delta_b)/h(\theta)$ versus the cell inclination $\theta$ for two different bubble diameters $d_b=22~$mm and $d_b=33~$mm (Cell 2). The sum of both lubrication films thickness  decreases by a factor 3 between $\theta=0^\circ$ and $80^\circ$, irrespective of the bubble size. Similar to the observations on the bubble speed, this drop of the normalised total lubrication film with $\theta$ is more significant at higher angles. Now, the effect of cell gap $h$ can be further explored if the same quantity is compared for two different cells (Cell 2 and 3) at a given bubble diameter. This is shown in Figure~\ref{fig:delta_rel_tot}(d) for $d_2=33~$mm. The lubrication films occupy a more important part of the cell for the larger gap ($\lozenge$, Cell 3, $h_0=5.21~$mm), reaching 30 \% even for the vertical case ($\theta=0^\circ$). This can be explained by the capillary number effect, as expected from the Bretherton scaling. Indeed, Ca$=\eta v_b/\gamma$ is larger for a larger gap since bubbles rise faster when cell gap is increased. As the cell is further inclined, the total film thickness drops faster in Cell 2 ($\vartriangleright$, $h_0=2.29~$mm) as compared to the one in Cell 3 ($\lozenge$, $h_0=5.21~$mm).

\section{Discussion}
\label{sec:discussion}

In this section, we discuss the possible origin of the departure of the measured bubble rise speed from the Taylor-Saffman limit (Eq.~\ref{eq:Velocity_Maxworthy}) as well as the loss of symmetry between the top and bottom lubrication films when the cell is inclined. Indeed, although the Taylor-Saffman speed accounts for the effective gravity by introducing $g \cos\theta$ instead of $g$ in Eq.~\ref{eq:Velocity_Maxworthy}, the buoyancy force perpendicular to the channel walls was not considered when they use the quasi-2D Hele-Shaw approximation. This buoyancy component is non-zero as soon as $\theta >0$ and increases when increasing $\theta$. 

\begin{figure}[h!]
\centering
      \includegraphics[]{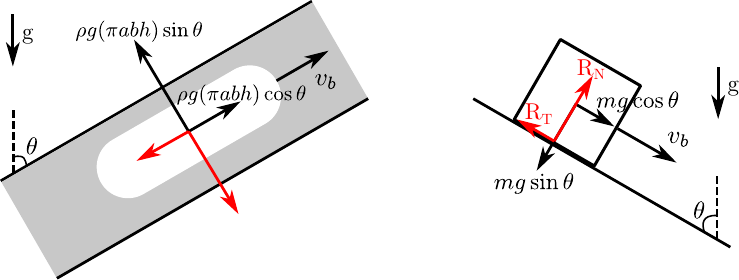}
  \caption{Sketch of the analogy between a rising bubble and a solid block sliding down a plate. The forces due to gravity are in black and while the ones from the surrounding media are in red.}
\label{fig:dessin_forces}
\end{figure}

We propose here a toy model based on the analogy between a bubble rising and a solid body sliding down a plane (Figure~\ref{fig:dessin_forces}). The additional power dissipation due to friction $P_f$ can be written as proportional to the bubble velocity $v_b$ and the force orthogonal to the plate, $P_f=\kappa (\rho \pi a b h \sin \theta)v_b$ where $\kappa$ is dimensionless and can be interpreted as the analogue of a friction coefficient. The injected power $P_b=\rho (\pi a b h \cos \theta) v_b$ is therefore dissipated not only by the classical term of bulk dissipation for a Poiseuille flow, leading to Eq.~\ref{eq:Velocity_Maxworthy}, but also by this additional friction $P_f$: \begin{equation}
    \rho (\pi a b h \cos \theta) v_b = \frac{12\eta v_b^2\pi b^2}{h}+ \kappa (\rho \pi a b h \sin \theta) v_b \, ,
\end{equation}
which can be rewritten as
\begin{equation}
    v_b = v_M (1-\kappa \tan \theta) \Leftrightarrow \alpha(\theta)=1-\kappa \tan \theta.
    \label{eq:modele_alpha}
\end{equation}
This phenomenological model agrees well with the experimental data (see dashed lines in Figure~\ref{fig:alpha_theta}), with $\kappa$ depending on the cell gap $h$. We find $\kappa=(6.5 \pm 0.5) \times 10^{-3}, (4.5 \pm 1.5) \times 10^{-3}$ and ($1.9 \pm 0.4) \times 10^{-3}$ for Cell 1, 2 and 3 respectively, meaning that $\kappa$ decreases when $h$ increases.  It is interesting to note that this model implies the existence of a critical angle $\theta_c=\text{arctan}(1/\kappa)$ beyond which the bubble should not rise anymore. Unfortunately, $\theta_c$ is close to 90$^{\circ}$ ($\theta_c>89^{\circ}$ for all cells presented in this work) and therefore cannot be reached experimentally.


The role of the component of the buoyancy force along the cell gap coordinate $y$ can then be interpreted as a pushing force on the bubble towards the upper plate of the cell (section~\ref{sec:lubrication}), analogue to the ground reaction for the solid body sliding down a plane (Figure~\ref{fig:alpha_theta}). The frictionnal losses are related first to the asymmetric local flow next to the bubble due to the top and bottom lubrication films; second to the matching between this local flow and the symmetric Poiseuille flow in the far field. However, this analogy cannot explain the observation that the top lubrication film  $\delta_t/h(\theta)$ scales with $\text{Ca}^{2/3}$ while the bottom film thickness does not present any Capillary number scaling. Note that these discussions are nevertheless heuristic and further theoretical investigation is needed to justify experimental observations.


\section{Conclusion}
In this paper, we investigated experimentally the time-averaged bubble speed and aspect ratio of freely rising bubbles in an inclined Hele-Shaw cell, as well as the associated lubrication films. Our experiments focus on low Reynolds numbers ($Re_{2h} \ll 1$), so that the dynamics is governed by viscous and pressure forces alone. We demonstrated that taking this problem as being the same as the vertical one with an effective gravity $g \rightarrow g \cos \theta$ is not adequate. Indeed, bubbles rise slower than expected with an effective gravity and the discrepancy grows with the angle $\theta$. We also showed that this difference depends on the cell gap while the liquid viscosity does not influence this effect. 

We attribute this decrease in bubble rise speed to the component of the buoyancy force along the cell gap, orthogonal to the cell walls. A bubble in an inclined Hele-Shaw cell experiences a force towards the top wall and hence, it does not occupy the channel gap as it would if the channel were vertical. In fact, by rewriting \citet{Taylor_1959}'s results based on the Hele-Shaw approximation in terms of the power balance between buoyancy and viscous dissipation, we suggest that the above rise speed discrepancy should arise from an additional term, analogue to the friction coefficient of a solid body down an inclined plane. This may result in a non-zero wall-normal component of velocity in the bubble's neighbourhood whose magnitude increases with the tilt angle $\theta$. Thereby, we propose that the rise velocity of a large elliptical bubble obtained with an effective gravity $g \cos \theta$ should be corrected by a factor $1-\kappa \tan \theta$, which fits well the experimental data. 

By measuring the lubrication films, we reported a strong difference between the film on top of and underneath a rising bubble. The latter becomes significantly larger than the former and this difference increases as the tilt angle is further increased towards $90^\circ$, leading to a bottom film thickness as much as four times larger than the top one. This phenomenon seems to be independent of the bubble diameter in our experiments. However, the asymmetry is stronger in a larger cell and so, there is a definite dependence on the channel gap size $h$. The top lubrication film follows the Bretherton law at low capillary number, $\delta_t/h \propto \text{Ca}^{2/3}$, over various tilt angles $\theta$.  But the proportionality factor depends on the cell gap and it is perhaps not as universal as suggested by previous results in vertical/horizontal Hele-Shaw cells, or cylindrical tubes. Moreover, our data for the bottom film thickness does not present such a power-law scaling. In summary, the liquid drainage from the top film and its relation to the buoyancy force along the wall-normal coordinate, i.e., along the cell gap, is crucial to the understanding of bubble dynamics and its associated lubrication films in an inclined thin channel. This symmetry breaking between the top and bottom lubrication films could be due to a flow generated by the bubble in the cell gap, which is a possible explanation of the factor $1-\kappa \tan \theta$ for the bubble velocity.
Further theoretical investigation would be required to understand this relation and shed light on the physical mechanisms driving the bubble velocity and lubrication films in inclined configurations.

%


\appendix

\section{Cell characteristics}
\label{sec:htheta}

The characteristics of the three cells used in this work (gap and plate thickness) have been carefully measured. We used a Chromatic Confocal Sensing (CCS) optical pen (STIL) to quantify either the thickness of the glass plate only, or the thickness of the glass plate and the gap when the pen is close enough to the cell plate. This technique enabled us to measure $h (\theta)$ locally along all the path that the rising bubbles take for Cell 2 and Cell 3. Unfortunately, this was not possible for Cell 1 because of the optical pen range of measurement. We have therefore used an additional technique, and determined the cell gap by filling the cell with a known quantity of water, which provides a global measurement of $h (\theta)$.

\begin{figure}[h!]
\centering
      \includegraphics[width=0.5 \linewidth]{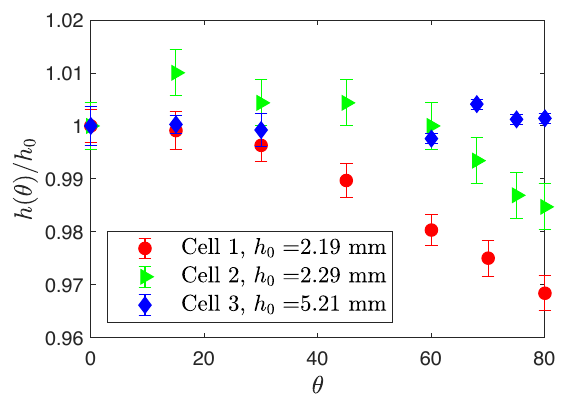}
  \caption{Evolution of the cell gap $h(\theta)$ normalised by $h_0=h(\theta=0^\circ)$ with the cell tilt angle $\theta$.}
\label{fig:h_Theta}
\end{figure}

The variation of the cell gap $h$ with the tilted angle $\theta$ is presented in figure~\ref{fig:h_Theta} for the three cells used in our experiments. The gap decreases as $\theta$ increases for Cell 1 and 2, while no noticeable variation is reported for Cell~3. The cell gap is imposed by spacers glued to the glass plates. When tilting the cell, due to gravity, the upper glass plate weights on the glue and tends to slightly squash and stretch it, resulting in a slight decrease of the spacing between both plates. When the inclination angle $\theta$ increases, this effect is expected to become stronger. In addition, plate mechanical bending may also play a role. This might explain why the most important gap variation is observed in the thinner cell (Cell 1), which has the lowest plate thickness ($e_g=3.9$~mm instead of 5~mm for cells 2 and 3, see Table~\ref{table:cells}). Although the cell gap variations only goes up to 3~\%, it is important to account for its precise value as the predicted bubble terminal velocity in the viscous regime (Eq.~\ref{eq:Velocity_Maxworthy}) is expected to vary as $h^2$ (see section~\ref{sec:expsetup}).

\section{Bottom lubrication films}
\label{sec:deltab}
\begin{figure}[t]
\centering
      \includegraphics[width=0.85 \linewidth]{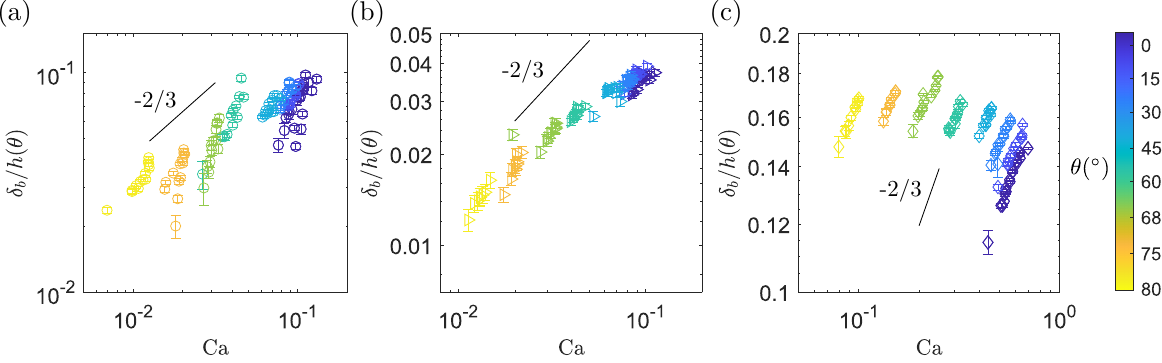}
  \caption{Non-dimensional bottom lubrication film thickness $\delta_b/h(\theta)$ as a function of the capillary number Ca$ = \eta v_b/\gamma$ for different inclination angles $\theta$ (see colorbar). \textbf{(a)} Cell 1 ($h_0=2.19~$mm), \textbf{(b)} Cell 2 ($h_0=2.29~$mm) and \textbf{(c)} Cell 3 ($h_0=5.21~$mm).}
\label{fig:deltab_Ca}
\end{figure}

Figure~\ref{fig:deltab_Ca} (a-c) shows the evolution of the normalised bottom lubrication film $\delta_b/h$ against the capillary number for Cells 1-3. Colors indicate the inclination angle $\theta$. Data from Cell 1 and Cell 2 present a strong scatter but, at a given angle, $\delta_b/h$ decreases monotonically with the Capillary number. The same trend is observed for the case of Cell 3. Note that the ordre of magnitude of $\delta_b/h$ is not the same between Cell 3 ($h_0=5.21~$mm) and the other two thinner cells. We further observe 2/3-scaling in Cell 3 such that $\delta_b/h=c(\theta) Ca^{2/3}$, where $c(\theta)$ differs for each inclination. However, no such scaling is discovered for data from Cell 1 and 2.

\end{document}